# Design of moveable and resizable graphics

We are communicating with computers on two different levels. On upper level we have a very flexible system of windows: we can move them, resize, overlap or put side by side. At any moment we decide what would be the best view and reorganize the whole view easily. Then we start any application, go to the inner level, and everything changes. Here we are stripped of all the flexibility and can work only inside the scenario, developed by the designer of the program. Interface will allow us to change some tiny details, but in general everything is fixed: graphics is neither moveable, nor resizable, and the same with controls. Author designed an extremely powerful mechanism of turning graphical objects and controls into moveable and resizable. This can not only significantly improve the existing applications, but this will bring the applications to another level. (To estimate the possible difference, try to imagine the Windows system without its flexibility and compare it with the current one.) This article explains in details the construction and use of moveable and resizable graphical objects.

## Introduction

When anyone switches on his PC, he usually sinks into the world of rectangular windows. We can easily move all these windows, resize them, overlap or put them side by side. At any moment we can reorganize the whole screen view to whatever we really need. It wasn't this way from the beginning of the computer era; it became the law after Windows conquered the world. This is an **axiom 1** in modern day programming design: <u>on the upper level all objects are moveable and resizable</u>. To make these features obvious, windows have title bars, to grab and move them, and borders, to resize them. Moveable and resizable are standard features of all the windows, and only for special purposes these features can be eliminated.

Usually the goal of switching on the computer is not to move some rectangular windows around the screen; we want to do a bit more, we start one or another application, step into inner level, and everything changes. Here, inside the programs, we are doing the real work we are interested in, and at the same time we are stripped of all the flexibility of the upper level, we can only do what the designer of the program allowed us to do. The design can be excellent or horrible, it can influence the effectiveness of our work in different ways, but still it is awkward that users are absolutely stripped from any control of the situation. Did you ever ask the question about the cause of this abrupt change? If you ever had, then you belong to the tiny percent of those, who did. And I would guess that the answer was: "Just because. These are the rules."

Unfortunately, these ARE the rules, but the rules are always based on something. The cause of this huge difference between levels that on the upper level we have only one type of objects – windows, and on the inner level there are two different types: controls, inheriting a lot from windows, and graphical objects that have no inheritance from them and are absolutely different. The addition of these graphical objects changes the whole inner world.

The inheritance of controls from windows is not always obvious, as controls often do not look like windows. Controls have no title bars, so there is no indication, that they can be moved; usually there are no such borders that inform about the possibility of resizing. But these inherited features of all controls (moveable and resizable) can be easily used by the programmers, for example, in the form of anchoring and docking. The most important thing is not *how* controls can be moved and resized, but that for them moving and resizing *can be organized* without problems.

Graphical objects are of absolutely different origin than controls and by default they are neither moveable, nor resizable. There are different ways to make things look not what they are in reality (the programmers are even paid for some knowledge of such tricks), and one of the often used techniques is the painting on top of the control: any panel is a control, so it is resizable by default, and with the help of anchoring / docking features it is fairly easy to make an impression as if you have a resizable graphics, which is changing according with the resizing of the form (dialog). By default panels have no visible borders, and if the back color of the panel is the same as of the parent form, then there is no way to distinguish between painting in the form or on the panel, which resides on it. Certainly, such "resizing" of graphics is very limited, but in some cases it is just enough; all depends on the purpose of application. Another solution for resizing of rectangular graphical objects is the use of bitmap operations, but in most cases it can't be used because of quality problems, especially, while enlarging the images. Both of these tricky solutions (painting on panel or using bitmap operations) have one common defect – they can be used only with the rectangular objects.

If any limited area is populated with two different types of tenants (in our case - controls and graphical objects), that prefer to live under different rules, then the only way to organize their peaceful residence and avoid any mess is to force them to live under ONE law. Because the currently used graphics is neither moveable, nor resizable, then the easiest solution is to ignore these controls' features, as if they don't exist. That is the reason, why the percent of applications, which are allowing users to move around any inner parts, is next to nothing. Thus we have **axiom 2**: <u>on the inner level the objects are usually neither moveable, nor resizable</u>. What is interesting, these two axioms bring us to the absolutely paradoxical situation:

- on the upper level, which is not so important for real work, any user has an absolute control of all the components, and any changes are done easily;

- on the inner level, which is much more important for any user, because the real tasks are solved here, user has nearly no control at all, and if he has, then it is very limited and is always organized indirectly through some additional windows or features.

Axioms I mentioned were never declared as axioms in strict mathematical way; at the same time I never saw, read or heard even about a single attempt to look at this awkward situation not as an axiom and to design any kind of application on different foundation. Programmers got these undeclared axioms from the main software developer (Microsoft) and are working under these rules for years without questioning them.

Certainly, anyone can easily remember one or another example of the graphical object, which was changing its sizes, for example, in Paint the dotted line is moving with your mouse. It is a fairly easy trick that can be done, for example, with the help of XOR operation and it has nothing to do with the real moving or resizing of the objects. Such type of "moving or resizing" is only an imitation, but in some situations it works.

The problem of design the moveable / resizable graphics is not a theoretical idea of "would be nice to have" type. For many years I was designing very complicated programs for engineering and scientific tasks in absolutely different areas: voice analysis and speech recognition, thermodynamics, telecommunication, analysis of electricity networks and some others. Though the aims of these programs had nothing in common, all of them required to high extent the use of different forms of plotting and the quality of users' analysis of the most difficult problems in each of these areas (and in many others) highly depends on the quality of graphical presentation of data and results. Because every user has his personal view on how the system must look to be the best instrument for his own work, the development of such systems is going in parallel with never ending discussions and even quarrels between designer and users. I think that anyone who is designing such type of applications is familiar with this situation and has to work under the same pressure of multiple requests which often can demand even the opposite things. For such complicated systems the use of moveable graphics will not only put the end of these endless discussions about the best view by giving users the full control of it, but this instrument will significantly increase the effectiveness of engineers' work with such applications, which from my point is the main goal. I saw again and again that graphics, designed and fixed by the developer of the application, became not only the main problem in further improvement of engineering and scientific software, but became the real barrier in exploration and solving of the most interesting problems.

Big engineering and scientific programs are brilliantly designed, but development of the big applications takes some time, so every user is restricted to whatever vision of the situation the designer had one, two or three years ago. These are the consequences of having non-moveable graphics – everyone is working with *designer-driven applications*.

"In science, finding the right formulation of a problem is often the key to solving it…" [1]. When I started to work on the problem of transferring unmovable graphics into moveable and resizable, I began with the analysis of their principal differences and features, that I would like to implement. The goal was not to make some kind of graphical objects moveable (for individual objects of particular type anything can be done), but to find the general solution. Though initially I was looking for the solution for scientific and engineering plotting, which usually has a rectangular shape, that was only some kind of experimental model. I was looking for the general solution and I found it. Before describing the whole algorithm I want to emphasize that:

1. The algorithm can be used in different areas and with arbitrary forms of objects.

2. The algorithm can be and must be looked on separately from the consequences of using it. Classes and algorithm for solution may be different, but if there is any form of moveable and resizable graphics, then it is the base for absolutely new paradigm – *user-driven applications*.

3. It is not simply an idea of "would be nice to have" type. The classes and algorithm of new plotting were designed and are working now; at the end of this article there are several links to applications and documents.

## Basic requirements

As a programmer and a designer of very complicated systems I would prefer to get this moveable graphics simply as one of the instruments from the package I am using for years – Visual Studio; unfortunately it isn't there (YET!), so let's write down the imaginable scenario of the best solution that somebody has already designed and handed to me. What I really need and would like to have.

1. I need an easy way to declare any object in the form (dialog) moveable and resizable.

2. Easy doesn't mean primitive. The configuration of my objects can be of any level of complexity, and changing of configuration may be influenced by a lot of different things. For example, some objects may allow any changes, others may need fixing of some parameters (sizes); parts of the objects may generate restrictions on other parts' changes. And still the whole variety of possible reconfigurations must be easy to understand and implement.

3. These features – moveable and resizable – must be added like an extra "invisible" feature, so it will be not destroying any image, but still it will be obvious that it is there and I can use it while working with the application. In some cases I can demand to show these features to me by some sort of additional visualization, but usually I would prefer to go on without any extra lines or marks.

4. Not only the new classes must be easily declared moveable and resizable, but the easiness must be also in adding these features to already existing classes; just touch them (or touch the keyboard several times), and they will change ([2]). These features must be additional; nothing of the case "all or none". Objects of the same classes can be used with or without these new features in order to start using them piece by piece and in doing it to see what they will bring into already existing complicated applications.

5. Using of these features must be as simple as with the windows on the upper level: press and move, press and reconfigure and even press and rotate, which is not organized for windows but can be extremely useful for many graphical objects. And if it is useful, it must be organized without any limitations, because we are talking about an imaginable perfect scenario.

Some of these "would like to have" look like conflicting (simple and with all the possibilities you can only imagine) and even alternative (not visible and obvious), but I am not writing the article about the future of programming in 2025. The biggest secret that all these things were designed and are working now; anyone can use it and move on with these results.

## The main idea - contour presentation

I am constantly working with C# language, so all the mentioned programs were done in this language, I am going to show the code samples in C# and use the terms from this language. But the algorithm and designed classes are not linked only to this language, they can be easily developed with other instruments; I simply prefer to use C#. Mostly the codes of the samples will be from the **Test_MoveGraphLibrary** project; all codes from this project are available (see link at the end).

There were two ways to add moveable / resizable features to the objects: either to use the idea of interface or of the abstract class; after trying both ways I stopped on an abstract class. Any object that would become moveable and resizable must be derived from the abstract class `GraphicalObject` that declares three crucial methods. (Closer to the end of this article, after describing all standard techniques and some special cases I'll write about the back door around this *must be*.)

```csharp
public abstract class GraphicalObject
{
    public abstract void DefineContour ();
    public abstract void Move (int cx, int cy);
    public abstract bool MoveContourPoint (int i, int cx, int cy, Point ptMouse,
                                           MouseButtons catcher);
    …
}
```

My idea of making graphical objects moveable and resizable is based on **contour presentation;** it is the core of the whole design, and any object that is going to be involved in moving and / or resizing must have a contour. For graphical objects the contour is organized in `DefineContour()` method.

The standard words from the graph theory often conflict with the meaning of the same words used by those who are programming under Windows for many years, so I'll try to use different words with obvious meanings.

Any graphical object that needs to be resizable and / or moveable will have a contour consisting of nodes and their connections, and these two parts of contours are used for different purposes. Contour is not duplicating the shape of the object; on rare occasions it can, but it would be really rare. Contour looks more like a skeleton, which allows the flesh around it to move as a single body, only I often use contour which is out of the painted object, and it would be difficult to imagine the skeleton being out of the flesh. In this case contour looks like a frame, but contour may consist of the disjoint sets, and still it will be the single contour. The possibilities look amazingly wide, but that is because contour was designed to cover any possible scenario of any real or imaginary object finding its way into programming world.

**Nodes** (class `ContourApex`) are used as sensitive areas which can be moved separately, thus providing reconfiguration or resizing of the object. Method `MoveContourPoint()` must include the code for individual movements of all the nodes that will be the subjects of such movements. Each node has its sensitive area, though for special cases, when this node must be excluded from the sole movement, this area will be null. When the node's area is not null and the mouse cursor is moving across this area, then the shape of the cursor can be changed to inform that by pressing the mouse button the node can be grabbed and moved. There are some common rules for conformity of possible movements of the object and the shape of the mouse cursor that signals about them, but there is some flexibility in defining this pair of things; I'll talk about them in describing the design of the real contour. Sizes and forms of the sensitive nodes can vary, and the simplicity of using this moveable graphics depends on the designer's decisions in organizing nodes.

**Connections** (class `ContourConnection`) are used for grabbing and moving the whole object; this movement is described in `Move()` method. Each connection has its own sensitive area in which the cursor's shape will be also changed to inform the user, that something underneath (the whole object!) can be grabbed and moved. Very often the area of graphical object is used for some mouse generated events that are aimed at starting different actions with the object or different changes. Implementation of the moving and resizing is based on mouse generated events in the areas of the nodes and sensitive areas around connections, thus an addition of contour may be the cause of the conflict between old commands and new requirements. To minimize this conflict the sensitive areas around connections usually are thin enough (their size can be even decided by the users as an additional parameter in application), though there are situations when the sensitive areas around connections are made as wide as possible, filling maximum of the object.

Nodes and connections, as they were described, can be the construction elements of the widest variety of contours; each design depends on the goal of particular graphical object. Let's look how all this works in the real sample.

## SimpleHouse – a classical moveable and resizable object

There was a nice time when you could take a pen and a sheet of paper and build your own house. No restrictions, just imagination. (And no bills, repairs, taxes – I want to be back!!!) If you liked your first house you could put another one near by, you could construct a street, a village, a town (if you had enough space on the paper…). There was only one dark side at that time: if you drew anything wrong, you had either to find the way to erase it, or you had to abandon the whole project and start the new one on the new sheet of paper. Not a bad idea, but it's a pity to abandon the nice village only because one cottage got the wrong color. So we are going to design moveable and resizable buildings, and all their colors will be changeable.

We are going to design the class `SimpleHouse`; this is part of the **Test_MoveGraphLibrary** project; you can find the code in **SimpleHouse.cs** file and objects of this class are used in **Form_Houses.cs** (menu position **Houses**). I call these houses simple, because the basic form of the house will be a rectangle plus a triangular roof (Figure 1); the majority of real houses falls under this description. Houses can be wide or narrow (at least two windows), high or low (at least one floor), the roof can be also high or low and its top can move to one side or another. I decided that the good place for the number will be somewhere on the roof; don't worry, the postman in our town will be Carlson from Astrid Lindgren; if anybody forgot – this nice guy can fly.

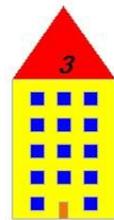

Fig. 1

To make our houses moveable and resizable we derive their class from `GraphicalObject` and we need to override its three abstract methods

```
public class SimpleHouse : GraphicalObject
{
    int nNumber;
    Font font;
    Rectangle rcHouse;
    Point ptTop;
    int roomSize;
    …
```

Before writing any code we have to make the <u>decision about the placement and type of the contour</u>. This is a real game for the kids, so nodes and connections must be absolutely obvious without any extra visualization. Corners of the house plus the point at the top of the roof look like the good places to press them and move, thus changing the sizes of the house, so these five points look like ideal places for the nodes. The borders of the rectangle plus the edges of the roof look like good places to grab and move the whole house to a new place (relocation is really easy), so these lines would be our connections. Usually there can be several different variants of contour for any object, and later I'll show such samples. In the case of `SimpleHouse` there also can be other variants, but if we decided about the nodes and their connections, then we can define the contour.

```
public override void DefineContour ()
{
    ContourApex [] ca = new ContourApex [5];
    ca [0] = new ContourApex (0, new Point (rcHouse .Left, rcHouse .Top),
                        new Size (0, 0), MovementFreedom .Any, Cursors .Hand);
    ca [1] = new ContourApex (1, new Point (rcHouse .Right, rcHouse .Top),
                        new Size (0, 0), MovementFreedom .Any, Cursors .Hand);
```

```
    ca [2] = new ContourApex (2, new Point (rcHouse .Right, rcHouse .Bottom),
                   new Size (0, 0), MovementFreedom .Any, Cursors .Hand);
    ca [3] = new ContourApex (3, new Point (rcHouse .Left, rcHouse .Bottom),
                   new Size (0, 0), MovementFreedom .Any, Cursors .Hand);
    ca [4] = new ContourApex (4, ptTop, new Size (0, 0),
                   MovementFreedom .Any, Cursors .Hand);
    ContourConnection [] cc = new ContourConnection [6] {
                new ContourConnection (0, 1), new ContourConnection (1, 2),
                new ContourConnection (2, 3), new ContourConnection (3, 0),
                new ContourConnection (0, 4), new ContourConnection (1, 4) };
    contour = new Contour (ca, cc);
}
```

Initialization of any new `ContourApex` is using five parameters:

`int nVal`                each node of the contour must have a unique number; the numbers must be from the range [0, nodes-1]; the order of numbers doesn't matter.  I began from the left top corner, went clockwise throughout the corners and then added the roof top.  All the numbers must be different, as later they are used for initialization of connections;

`Point ptReal`            a real point on the screen; in our sample it is either a corner of the house or the roof top;

`Size szRealToSense`      this is a shift from the real point to the middle of the node, associated with this point; in the case of our houses we don't need to move the nodes anywhere from the real points, because these nodes will be not shown at all, but there are situations when it is very helpful to move the node slightly aside and not close the real image of the object; such decision of moving nodes aside from real points is used, for example, in case of very complicated and informative engineering plotting;

`MovementFreedom mvt`     possible sole movements of the node;

`Cursor cursorShape`      cursor shape above the node.

Regardless of what you put for the particular `ContourApex` as the fourth parameter, all nodes will move around when the whole object is moved; `MovementFreedom` parameter describes only the opportunities for separate movings of this node when you want it (or not) to paricipate in reconfiguration.  The possible values are

```
enum MovementFreedom { None,       node is not used for reconfiguration
                       NS,         can move only Up or Down
                       WE,         can move only Left or Right
                       Any };      can move in all directions
```

The last parameter in the initialization of `ContourApex` defines the shape of the mouse cursor when the mouse is above the node.  There are standard expectations that when, for example, cursor has the form of `Cursors.SizeWE` then the object underneath can be moved only left or right, so it's better not to organize any confusions by setting a very strange pair of the last two parameters.  At the same time I didn't want to put strict rules on the cursor's shape based on possible movements, so you have this fifth parameter for your decision.

The type of contour, we have for `SimpleHouse`, is the most common case of the contours; it has several nodes and several lengthy enough connections between them.  There can be special and very useful types of contours, which differ from this standard one, and I'll describe them in details later, when we'll finish with the `SimpleHouse`.

**An important remark**.  Though the whole idea of moveable and resizable graphics is based on contour presentation, this method is usually the only place where you will have to think about the contour!  The rule is: organize the contour and forget about it.  There is one exception of this rule; I'll write about it a bit further when we'll look into individual movements of the nodes.

`Move(cx, cy)` is the method of forward <u>moving the whole object</u> for the number of pixels, passed as the parameters.

Drawing of any graphical objects with any level of complexity is usually based on one or few very simple elements (`Point` and `Rectangle`) and some additional parameters (sizes).  While moving the whole object we are not changing the sizes, so we have to change only positions of these basic elements.  In the case of the `SimpleHouse` we have two such elements: `Rectangle rcHouse` and `Point ptTop`.

```
public override void Move (int cx, int cy)
{
```

```
    rcHouse .X += cx;
    rcHouse .Y += cy;
    ptTop += new Size (cx, cy);
}
```

MoveContourPoint(i, cx, cy, ptMouse, catcher) is the method of <u>moving the individual node</u>.  Method is returning the `bool` value, indicating if the required movement is allowed or not; in case of forward movement the `true` value must be returned in case any of the proposed movements along X or Y scales is allowed.  If the movement of one node is resulting in synchronous relocation of a lot of other nodes, then it is easier to put inside this method the call to `DefineContour()`, and then it doesn't matter what value is returned from `MoveContourPoint()`.  This may happen for forward movement but always happens with rotation when all nodes must be relocated.  This is the exception of the previously mentioned rule, that contour is not even thought about anywhere outside the method of its definition.

Method `MoveContourPoint()` has five parameters, but not all of them are used each time

| | |
|---|---|
| `int i` | identification number of the node – the same number that was used in `DefineContour()` method to identify this node; if the node is not involved in seperate movements, then you don't need to mention it in `MoveContourPoint()` method; |
| `int cx, int cy` | movements (in pixels) along two scales; positive numbers for moving from left to right and from top to bottom; use these parameters if you are writing the code for forward movements; |
| `Point ptMouse` | the positiion of mouse cursor; for calculations of the forward movements I simply ignore this parameter and use the previous pair; for calculations of rotation I ignore the previous pair and use only this cursor's position; I found it much more reliable for organizing any rotations; |
| `MouseButtons catcher` | informs, which mouse button was used to grab the object; if by the logic of application the object can be grabbed by any mouse button, then simply ignore this parameter; if the move is allowed only by one button, then it is a useful parameter; in case when the node can be involved in two types of movement (forward movement and rotation) this is an extremely useful parameter to distinguish between them. |

Usually the method of individual nodes' moving will be the longest of all three methods; however, there are interesting situations when this method will consist exactly of one line.  Because the method must include the code for moving of each involved node, it can be a long one, but it will be not complicated, as more than often the code for different nodes can be partly the same.  For `SimpleHouse` I will comment here not the full method but parts of the code for two different nodes: the left top corner and the roof top.  In the whole method and can see that the code for two corners on each side of the house is partly the same.

```
public override bool MoveContourPoint (int i, int cx, int cy, Point ptM,
                                       MouseButtons catcher)
{
    bool bRet = false;
    if (catcher == MouseButtons .Left)     // resizing only with the left button
    {
        if (i == 0)      // Left-Top corner
        {
            if (rcHouse .Height - cy >= minHeight)    // compare with minumum height
            {
                rcHouse .Y += cy;
                rcHouse .Height -= cy;
                ptTop .Y += cy;     // roof height is not changing
                bRet = true;
            }
            if (rcHouse .Width - cx >= minWidth)    // compare with minumum width
            {
                rcHouse .X += cx;
                rcHouse .Width -= cx;
                ptTop .X += cx;
                ptTop .X = Math .Min (Math .Max (rcHouse .Left + minRoofSide,
                                        ptTop .X), rcHouse .Right - minRoofSide);
```

```
            bRet = true;
        }
    }
```

As I mentioned before, here you are not doing anything with the contour itself, but only checking the possibility of proposed movements of the real points associated with the nodes.  For the left top corner of the house I am:

- comparing the proposed new height of the house with the minimum allowed height; if the top of the house is moving, then I also have to move the roof top;
- comparing the proposed new width of the house with the minimum allowed width; if the width of the house is going to be changed, then I have to determine the new position of the roof top; I am trying to keep it in relatively same position, but the roof top can not be closer to any side than the predefined minRoofSide;

The piece of code for the roof top (identification number 4) is similar: I have to compare the proposed new roof top position with the minimum allowed roof height and with the minimum allowed distances to the sides of the house.

```
        else if (i == 4)      // Roof top
        {
            if (ptTop .Y + cy <= rcHouse .Top - minRoofH)    // compare with minumum
            {
                ptTop .Y += cy;
                bRet = true;
            }
            if (rcHouse .Left + minRoofSide <= ptTop .X + cx  &&
                ptTop .X + cx <= rcHouse .Right - minRoofSide)
            {
                ptTop .X += cx;
                bRet = true;
            }
        }
```

This was the typical case of organizing moveable and resizable graphical object.  We designed the contour of several nodes and their connections; we developed methods for moving the whole object and resizing it.  The `SimpleHouse` objects now can be moved and resized; lets' organize the whole process.

## And the boss is Mover

The easiest way to move any object around the screen would be to press it with the mouse, move it to the new position and release, so for the whole process only three mouse events are used: `MouseDown`, `MouseMove` and `MouseUp`.

From users' point of view the whole process must be simple: press, move (or rotate) and release.  And this is the only way users are going to tolerate this process; any additional complication will be not acceptable.  The designer that is planning to include moveable / resizable objects into his application certainly understands that somewhere and somehow all the information about the involved objects, positions of the nodes and connections, sensitive areas, their overlapping, order of drawing, type of movement and a lot of other additional things – all these things must be stored somewhere, analyzed and processed.  At the beginning I formulated the basic requirements to the whole moving / resizing process and wished, that there would be a class (better in Visual Studio, but unfortunately it's not there) that will provide all the desired features.  Such class exists, it is ready to do everything, and its only requirement is that the objects, involved in moving and resizing, were either derived from `GraphicalObject` or be a control (I'll write about this case later).

```
public class Mover
```

The name declares only the part of what this class is capable of; in addition to moving it is also doing all resizing and giving information about all possible things, associated with the process.  Class `Mover` is described in **MoveGraphLibrary_Classes.doc**, you can certainly use all its properties and methods, only one small advice, based on my experience (I am using the same class `Mover` for all my applications), but mostly on common sense.

If somebody will decide to move a building from one piece of property to another and to put it there in the original form, then chances are high that he will make some agreement with moving / construction company to do the job.  If all pieces were carefully marked, the house was disassembled and all parts were only waiting for some truck for relocation; if at that moment the owner would decide to take out some pieces to repair the deck somewhere else; well, as an owner of all and each part he certainly has the right to do it, but after this I wouldn't bet on his chances to see the building reassembled at the new place in the proper way.  I myself let the `Mover` always do the job and never try to help him, and because of this up till now I never had any problems.

Now to our real town from the file **Form_Houses.cs**.

I want all the houses in the town to be moveable and resizable, so I designed them as
```
public class SimpleHouse : GraphicalObject
```

Next I am declaring the company responsible for all movements and reconfiguring of the houses and also starting the construction of the town
```
Mover Movers = new Mover ();
List<SimpleHouse> Town = new List<SimpleHouse> ();
```

The new house will pop-up in the town at the same moment you'll click the *New house* button
```
private void Click_btnNewHouse (object sender, EventArgs e)
{
    SimpleHouse newhouse = new SimpleHouse (nNew, …, rc);  // initialize new house
    Town .Insert (0, newhouse);     // insert the new house into the List<>
    Movers .Add (newhouse);     // register the new house as moveable
    Invalidate ();
}
```

I slightly shortened the real code to show here only the lines that are really significant:

- Initialize the new house.

- Insert the new house into the official town's list of buildings.  City is responsible for painting of all the houses (look into `Paint()`), but I want every new house to be shown on top of all previous, so I always insert the new house on zero position and the painting is going from the end.

- Register the new house as moveable (and also resizable because of the type of contour it has).  This is the mandatory thing if you want the house to become moveable / resizable!  Contour itself gives an object the ability to be involved into moving / resizing process; this line adds the object to the set of moveable / resizable objects.
  ```
  Movers .Add (newhouse);
  ```

Several important things about Mover

1. `Mover` is working only with the objects that he was asked to take care of.  `Mover` has his own List of these objects; you, as a programmer, have an access to the elements of this list via standard indexing or you can use standard List<> methods (look into **MoveGraphLibrary_Classes.doc** for the description of implemented methods).

2. `Mover` doesn't know anything about the real objects; `Mover` is working only with the contours of the objects, which were included into his list.  It is the programmer's responsibility to make changes in parallel in the outer List (there can be several lists, or arrays or anything else) and the List inside `Mover`.  The most awkward situation would be if `Mover` would try to take care of the object which you have already deleted from the real world (from the form).  Don't blame anyone if you are doing your best to organize a mess, and anyway, I think that from time to time even programmers have to do something.  Not all objects on the form must be declared moveable / resizable, that is the programmer's decision to give this feature to all of them, some of them or none of them, so the programmer has to check that `Mover` is working on a correct List.  There are all methods to do it easily.

The real `Mover`'s power is shown in the code of three mouse events.  On **MouseDown** event I try to grab an object
```
private void OnMouseDown (object sender, MouseEventArgs mea)
{
    if (mea .Button == MouseButtons .Left)
    {
        Movers .CatchMover (mea .Location);    // start moving / resizing
    }
    else if (mea .Button == MouseButtons .Right)
    {
        …
        ContextMenuStrip = contextMenuOnHouse;
    }
    if (!Movers .MoverCaught)
    {
        BringToTop (mea .Location);    // if clicked not for moving, bring to top
    }
}
```

It doesn't matter how many moveable / resizable objects are there in the form; you need only one line to start moving / resizing process for any object

```
Movers .CatchMover (mea .Location)
```

This is the only piece of code in `OnMouseDown()` method that is really important for starting of any moving / resizing process; all other lines of code in this method are simply doing some helpful, but additional things, like opening the context menu or bringing to the top the house, that was clicked with a mouse.

**MouseUp** event will finish any moving / resizing process and release any previously grabbed object. The code is always extremely short.

```csharp
private void OnMouseUp (object sender, MouseEventArgs e)
{
    Movers .ReleaseMover ();
}
```

The third event - **MouseMove** - has to do everything on real moving / resizing, but the code is not complicated even here

```csharp
private void OnMouseMove (object sender, MouseEventArgs mea)
{
    Cursor cursor = Cursors .Default;
    foreach (SimpleHouse house in Town)
    {
        if (house .Inside (mea .Location))
        {
            Cursor .Current = Cursors .Hand;
            break;
        }
    }
    Movers .MovingMover (mea .Location);    // real moving / resizing
    if (Movers .MoverCaught)
    {
        Invalidate ();
    }
}
```

The first half of this method – the bigger part – has nothing to do with moving, and only changes the cursor's shape, if the mouse is over any house. The real moving again is done in one line of code. There is also the checking of `Movers` situation and if the `Mover` will signal that at this moment some object from his list was caught, then he will trigger the repainting. To avoid flickering of the screen don't forget to switch ON double-buffering in the form; it has nothing to do with the design of moveable / resizable graphics, but it is simply a nice feature from Visual Studio.

And that is all: we have our town of moveable and resizable houses. There are some additional and very useful things in the real application, like possible change of all the colors, saving the image into clipboard for printing, saving the picture into file and restoring it from file. Well, we prefer to live in a nice town, so we have all these nice features, but the main thing of organizing the `SimpleHouse` as moveable / resizable object is done and working.

`SimpleHouse` has a classical contour of several nodes and lengthy enough connections between them, but there can be other very interesting and in some cases very useful situations. Let's look at some of them.

## Special types of contours

### Case A. Same nodes – different connections.

Let's look at the case of the scale, which we need to make resizable only horizontally. Also let's decide first that the contour must be out of the scale's image. In need of absolutely nonresizable contour it would be enough to have four nodes close to the corners of the object, but because of the request for horizontal resizing there would be two additional nodes on the left and right sides, and these two nodes will differ from other four.

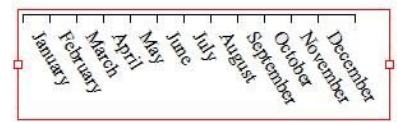

Figure 2. Resizable scale

Let's decide that we have such variables
```
int cxL, cxR, cyT, cyB      all four borders of the scale's rectangular area
int shift                   the distance of moving the nodes outside the scale's rectangle
```
The array of nodes can be organized in such a way
```
apexes = new ContourApex [6] {
```

```
            new ContourApex (0, new Point (cxL, cyT), new Size (-shift, -shift),
                             MovementFreedom.None, Cursors.SizeAll),
            new ContourApex (1, new Point (cxR, cyT), new Size ( shift, -shift),
                             MovementFreedom.None, Cursors.SizeAll),
            new ContourApex (2, new Point (cxR, cyM), new Size ( shift, 0),
                             MovementFreedom.WE, Cursors.SizeWE),
            new ContourApex (3, new Point (cxR, cyB), new Size ( shift, shift),
                             MovementFreedom.None, Cursors.SizeAll),
            new ContourApex (4, new Point (cxL, cyB), new Size (-shift, shift),
                             MovementFreedom.None, Cursors.SizeAll),
            new ContourApex (5, new Point (cxL, cyM), new Size (-shift, 0),
                             MovementFreedom.WE, Cursors.SizeWE)
};
```

The nodes were numbered from the left top corner going clockwise. For nodes with `MovementFreedom.None` (these are four nodes in the corners) the last parameter (cursor's shape) doesn't matter, and while showing the contour these nodes will be not shown at all, as if they do not exist (look at the figure). Still we need these nodes as the connections can go only between the nodes, and if we want to have the contour around the object, then we have to put all six nodes.

Contour's initialization is based on the array of nodes
`contour = new Contour (apexes);`

You may be surprised to see that while organizing this contour I didn't organize the array of connections, but used the `Contour` constructor with one parameter. In such case the array of connections is organized automatically by linking each node in array with the next one and then linking the last node with the first.

If we want to have the same horizontally resizable scale, but without the requirement for the contour to be out of the scale, then we can construct another contour, even using the same two small nodes in the middles of the sides: the single connection between them can go right through the middle of the scale.

```
apexes = new ContourApex [2] {
            new ContourApex (0, new Point (cxR, cyM), new Size ( shift, 0),
                             MovementFreedom.WE, Cursors.SizeWE),
            new ContourApex (1, new Point (cxL, cyM), new Size (-shift, 0),
                             MovementFreedom.WE, Cursors.SizeWE)
};
```

From the point of resizing the object these two solutions are absolutely identical: there are two small nodes on the sides; any node can be grabbed and moved left or right. The code for `Move()` method will be the same, as it is the same object. `MoveContourPoint()` method will be also the same (only different cases for the first parameter – identification number), as the code must be written only for the nodes, which are really involved in resizing.

The difference between two contours will be in the process of moving the whole object. In the first case (contour around the object) the scale can be grabbed for moving only in the areas close to the border of the object. In the second case (contour through the middle of the scale) the area for grabbing the object will be a thin horizontal strip somewhere in the middle. I am writing *somewhere,* because based on the changing texts and possibility of using different fonts this strip will change its place, and it would be difficult to find these nodes without special visualization. In the case of a single connection between two nodes the better solution would be to place the nodes at the ends of the main line, then at least there would be no problems in finding them.

In case when the scale's main line is always shown I would say that it would be the best solution to put the nodes at the ends of the main line, but… In many cases the scale can be shown without the main line: just ticks and texts or only part of them; for any user to find the nodes at the ends of the invisible line would be really a tricky thing. What I want to emphasize that there are always different ways to organize contours; while doing it you have first to decide what would be the best and most obvious solution for the users, who are going to work with such objects.

Both samples in this case still belong to the most common case of contours with several small nodes and connections between them. There can be other special cases of contours, and their special design will make some impact on the code of `MoveContourPoint()` method.

**Case B.    Moveable, but not resizable.**

Moveable, but not resizable objects are used very often, for example, in logical games. What is interesting, that the described system of nodes and connections can produce different contours for such objects, all of them will work perfectly, so it is simply for the designer to choose one of the solutions. Let's look at the square shape objects; object is described in **TwoNodesSquare.cs** file and the use of these objects is in **Form_ColoredSquares.cs**

As you can assume from the name of the file (TwoNodesSquare) the contour will be based on two nodes. They are standard small nodes; but here you can forget about the sizes of the nodes(!), because they will be not used at all; the nodes will be declared as not used for sole movements, so they will be automatically set to null.

```
public override void DefineContour ()
{
    Point ptM = Auxi_Geometry .Middle (rc);   // middle point of Rectangle
    ContourApex [] apexes = new ContourApex [2] {
        new ContourApex (0, new Point (ptM .X - 1, ptM .Y), new Size (0, 0),
                         MovementFreedom.None, Cursors.SizeAll),
        new ContourApex (1, new Point (ptM .X + 1, ptM .Y), new Size (0, 0),
                         MovementFreedom.None, Cursors.SizeAll)
    };
    contour = new Contour (apexes);
    contour .LineSensitivity = Math .Min (rc .Width, rc .Height) / 2 - 1;
}
public override void Move (int cx, int cy)
{
    ptLT += new Size (cx, cy);
}
public override bool MoveContourPoint (    )
{
    return (false);
}
```

These two nodes are placed in the middle of the square next to each other. Both of the nodes are not moveable, so the third function simply returns `false` (I mentioned previously that on some occasions this method can consist of one line). The connection between two nodes is very short (in this case – 2 pixels), but I increased the sensitivity of this connection up to the sides of the square, so the sensitive area of the connection will be very close to the circle inscribed into square, thus it will cover nearly 80 percent of the square. I doubt that any user of such program will pay any attention that it is impossible to grab the square in the small areas close to the corners; anywhere else you can press the mouse button and move the object.

Later I added to the same object another variant of the same type of contour (null nodes, increased sensitivity of connections) but with four nodes instead of two. All four nodes are still unmoveable separately, they stay far away from each other, the sensitivity of each connection is only half as much as in the previous case, but the nodes are located in such a way - each is half way from the corner to the middle of the square - that the combined sensitive area is covering 95 percent of the whole square. In **Form_ColoredSquares.cs** you can add to the view the new objects of `TwoNodesSquare` type, and the number of contour's nodes (two or four) will be one of the parameters while initializing the new object. This is a very interesting feature: objects of the same class may have different types of contours! This number of nodes is marked on the colored square while it is painted, so for each of the squares you'll know the number of nodes without trying to remember them. You can define the same size squares with different types of contours and compare for yourself if there will be a big difference in moving around the screen such graphical objects.

**Case C.  Another moveable, but not resizable.**

There is one more type of contour design for moveable, but not resizable squares. This solution may be funnier, but it will cover the whole square.

```
public override void DefineContour ()
{
    ContourApex [] ca = new ContourApex [1];
    ca [0] = new ContourApex (0, ptCenter, new Size (0, 0), MovementFreedom .Any,
                              Cursors .Hand);
    ca [0] .SenseAreaSize = size;     // change the node's size
    contour = new Contour (ca, null);
}
public override void Move (int cx, int cy)
{
    ptCenter += new Size (cx, cy);
}
public override bool MoveContourPoint (int i, int cx, int cy, Point ptMouse,
                                       MouseButtons catcher)
{
```

```
        bool bRet = false;
        if (catcher == MouseButtons .Left)
        {
            Move (cx, cy);
            bRet = true;
        }
        return (bRet);
    }
```

Here we have the unique contour consisting of a single node; certainly, there are no connections, but the sensitive area (the size) of this single node is changed in such a way as to cover the whole square. Because there are no connections, the movement of the object is based on the move of this single node, and if it is allowed, then `Move()` is called from inside `MoveContourPoint()` method. The node's sensitive area covers 100 precent of the square, so you can L_Press in any point to start moving.

Two last cases are for moveable and not resizable square objects. For the user the moving of these objects will look exactly the same: press – move – release; from the designer each of these special contours required specific code for `MoveContourPoint()` method.

**Case D.  Different shape of nodes.**

To make this system of nodes and connections even more flexible the shape of any node can be changed from square to circle

```
public override void DefineContour ()
{
    ContourApex [] ca = new ContourApex [1];
    ca [0] = new ContourApex (0, ptC, new Size (0, 0), MovementFreedom .Any,
                              Cursors .Hand);
    ca [0] .SenseAreaSize = nRadius;
    ca [0] .SenseAreaForm = ContourApexForm .Circle;   // change the node's shape
    contour = new Contour (ca, null);
}
```

Such contours, consisting of the single circle node, are used in **RegularPolygon.cs** file and in **Form_RegularPolygons.cs** you can see how it is working. In case B we had an example of an object that could receive different types of contours; here we have an example of different objects that have the same contour. To be correct: from programmer's point of view all these regular polygons are objects of the same type (same class), but for user they are absolutely different (they look different).

## Contours' summary
- Contour can consist of any number of nodes and their connections.
- Minimum number of nodes is 1. Contour may consist of a single node without any connections.
- Connections may exist only between nodes; on both ends of any connection there must be a node.
- Nodes can be moved individually thus allowing to reconfigure the object.
- By grabbing any connection the whole object can be moved.
- Contour may consist of a serious of connections between empty nodes. This makes an object movable, but not resizable.
- Each of the nodes has its own parameters, and by connecting nodes of different types it is easy to allow, for example, resizing along one direction but prohibit it along another. This means organizing a limited reconfiguring.
- It doesn't matter that some contours may represent graphical objects and others controls or groups of controls (I'll describe it later). All contours are treated in the same way, thus allowing the user to change easily the inner view of any application.

**Résumé**. With the combinations of nodes' shapes, nodes' sizes and the sizes of sensitive areas around connections any required contour can be organized.

## Visualization of contours

Previously I described how do design contours for the wide variety of graphical objects. Now any object in a program can be turned into moveable and resizable, and to get the best results from these new features users have to know:
- that any object inside application is moveable and resizable;
- how to use (to start) these features.

The first problem is more psychological than technical. How can engineers and scientists imagine that everything in my application **TuneableGraphics** is moveable and resizable, if they had never seen a single scientific program with moveable plotting? Lewis Carroll's solution of the problem with information simply written on top – *Eat me* – is definitely the best, but not for the programming world. In the world of applications may be the best solution would be to rely on human nature. So in **TuneableGraphics** application I put a small button in the left top corner of each form; I hope that people will click this button even of single curiosity, then they will see contours, will try to find what are these lines for, and everything else will become clear.

If users are familiar with the idea of moveable and resizable objects and are expecting all objects to have such features, then they will be looking for the most likely places for nodes and connections. Users will not even know such terms, they will be making the decision about the most likely places for moving and reconfiguring objects, and the designers' responsibility is to organize contour according with such expectations. Still, the easy way of switching contours ON / OFF will be useful in nearly any situation, so may be it is not bad to have that tiny button in the corner.

Requirement for good contours' visualization in any possible situation is a problem by itself. For solving it I set some default parameters that will give good results in a majority of situations and added some methods for parameters' changing.

Contours consist of nodes and their connections. By default:
- node is a small square with the side equal to 6 pixels;
- contour is shown in Red color; this will be the color of connections and borders of all the nodes;
- inner areas of all nodes are filled in White.

What were the ideas behind these default parameters? Contours are very helpful, but they are instruments; instruments must be obvious, but they can't be the main part of the view. Six pixel square is small enough not to disturb any image, but big enough to grab this mark by the mouse without any problems. Big images are rarely drawn in Red, so this color will be well visible in the majority of situations. Filling the inner part of the node with white color plus the red border of this tiny area informs the user about this additional but artificial thing; contours are doing their work perfectly, but the programmer must give users an easy instrument (one click) to switch the contours ON and OFF.

All the default contour's settings can be changed. I have already mentioned in the previous part how to change the shape of the node from square to circle. If you don't want to fill the inside of the node with White, you have to switch the flag into `false`:

```
ContourApex [] ca = new ContourApex [1];
ca [0] = new ContourApex (0, …);
ca [0] . SenseAreaClearance = false;
```

There is also an easy way to change the color of contour, but I'll show how to do it later. Now from the question of changing the contour's view let's turn to the code for painting contours.

The decision about what to draw and in which order is usually done inside the `Paint` event of the form. Here is the code from one form in **TuneableGraphics** application which shows very interesting graphical object – `Skyscrapers`, and there is also the view of this form which will make the explanation much more obvious.

```
private void OnPaint (object sender, PaintEventArgs e)
{
    Graphics grfx = e .Graphics;
    if (bShowAxes)
    {
        xyzCoor .Draw (grfx);     // coordinates
    }
    skyscr .Draw (grfx, xyzCoor); // Skyscrapers
    if (bShowContours)
    {
        Movers[0].DrawContour (grfx); // contour
    }
}
```

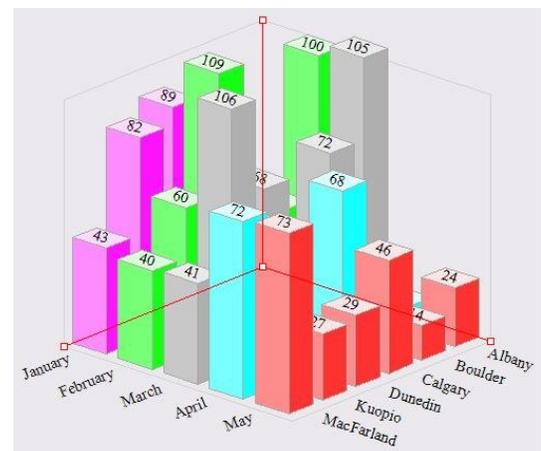

Figure 3. `Skyscrapers` and contour

There are two major graphical objects in this view. `Skyscrapers`, which are used for financial analysis, do not belong to moveable and resizable graphical types about which I am writing here; the object of this class simply resides under the jurisdiction of `XYZcoordinates` and is doing whatever it is told to do by the last one. (Class `XYZcoordinates` is included into **MoveGraphLibrary.dll**, so you can write your own classes and receive all the benefits of `XYZcoordinates` by using them together.) `XYZcoordinates` is a class of moveable / resizable objects

```
public class XYZcoordinates : GraphicalObject
```

As a class derived from `GraphicalObject` the `XYZcoordinates` object has all the features described before and you can see its contour on the picture; the contour of coordinate system copies the axes; the nodes are at the ends of the axes and in the crossing of coordinates. With the three nodes at the ends of axes you can change all the sizes of this picture. By the nature of this plotting the towers in front will often close from view the towers behind; by moving around the single node at the crossing of axes you can find in an instant the best view for any data set. Excellent, but that is where the problem can occur.

Contour belongs to `XYZcoordinates`, but it is never shown automatically together with the object to which it belongs (in this case xyzCoor). If it was drawn automatically with its "parent" object, then the picture of `Skyscrapers` will close it from view, and it would be difficult to locate that very useful node. It is still working in the same way either it is in view or not, and the changing of mouse cursor will signal you that you have at last located it, can press the mouse button and rotate the whole view, but still it will be a bit tricky to locate this tiny node behind the towers.

This is a very important feature of showing contours: as a programmer you have not only to decide about the queue of any objects you want to draw in your form, but you also have to decide about the order of contours in the same queue, and more often than not the contours will be in this queue not next to their "parent" objects. Even more: the "parent" objects can be excluded from drawing (by switching OFF one parameter in the tuning form you can take axes out of the above picture, and this will not change anything else), but still you have to decide about the moment when it would be the best to show contours. So in the code for drawing this picture you can see that first the coordinate system was drawn (there can be very different views of this system), then `Skyscrapers` and only after it the contour.

On the last picture you can see the contour, painted on top of everything else; if you look at the last piece of code you'll see that the object, responsible for drawing of the contour, is not the owner of the contour (object, derived from `GraphicalObject`), but `Mover`. At first it may look very strange, but here is the explanation of this situation.

Any object, derived from `GraphicalObject`, has a contour, but this object may be included into the list of moveable objects, or may be not. It would be a very common situation, when you would like to have in your form some nonmoveable objects and others – moveable. For example, you can put somewhere at the side of the form the small objects of the types (classes), that can be added as moveable, but these samples would be unmoveable. By clicking any of these samples you start the process of adding identical moveable object, but the sample itself will stay in the same position all the time. `Mover` has the list of all moveable objects of the form (of those that were included into **his** list); the object itself doesn't know either it was included into the list of moveable elements or not. `Mover` is the only one who knows about all and each contour that must be painted.

If you switch ON the contours in the main form of **Test_MoveGraphLibrary** application or play with different interesting forms of **TuneableGraphics** application, you will see everywhere that all contours are shown in one color. There is no problem in giving each contour its own color; in reality they have their own colors, only because they are too lazy to draw themselves and delegated this painting to `Mover` they are all shown in the same color. This color can be changed through one of `Mover`'s methods. (The words *lazy*, *painting* and some *brains*, definitely used to design this moveable / resizable graphics, reminded me of a famous process, which Tom Sawyer organized on one beautiful sunny day. Where is my dead rat on a string?)

## The colored doughnuts

Up till this moment I tried to illustrate everything with very simple, pure samples that were especially designed as simple as possible to explain the whole mechanism step by step. (`Skyscrapers` is an exception, but it came from another application.) In reality you would like to add contours to more complicated objects and may be to use with them both forward movement and rotation, so let's look at more complicated and real sample. In one of the descriptions I saw a picture of several coaxial rings, colored by sectors. I assume that such type of plotting is used somewhere in financial area; for me it became another exercise to check if I have missed anything in the design of moveable / resizable graphics or not. Class `DoughnutSet` is described in file **DoughnutSet.cs;** this picture and all samples of code for working with this class are from **Form_DoughnutSet.cs**.

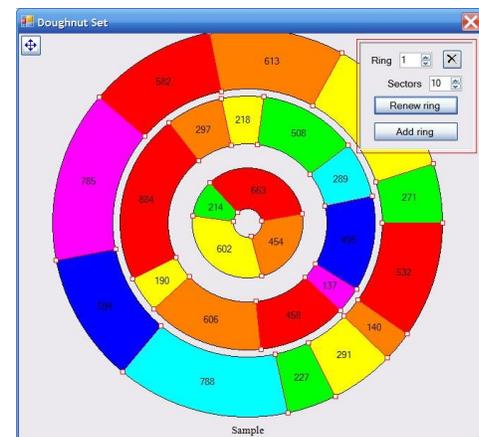

Figure 4. DoughnutSet

Each ring (class `DoughnutRing` in the file **DoughnutRing.cs**) consists of the number of sectors (minimum 2). To distinguish the sectors, each of them has its own color. Values can be shown as numbers or as percents, colors of sectors can be changed, colors of the texts for each sector can be changed, color of all borders can be changed, font to show the texts can be changed… All viewing parameters can be changed, only I took out this parameters' changing because it has nothing to do with the explanations in this article. You are free to add these features, but for the purpose of this application parameters are mostly selected at random.

```
public class DoughnutSet : GraphicalObject
{
    Point ptCenter;
    List<DoughnutRing> rings = new List<DoughnutRing> ();
    Title title;
    ….
```

Nodes are located in the points, where the sectors' borders cross with the inner or outer border of each ring (you can see these nodes even on a small picture). Pair of nodes on each border is connected to each other, but there are no connections between the pairs, so contour consists of unlimited number of disjoint sets. (You can see the best illustration of this strange contour in the main form of the application. Switch ON the contours by the small button in the corner and then move the picture of the `DoughnutSet` to the same location where you can see Y(x) plotting. The picture of `DoughnutSet` will be closed from view by that plotting, but its contour will be in perfect view above it.) L_Press on any node of `DoughnutSet` will allow moving it along radius, thus changing the width of the ring. R_Press on any node regardless of belonging to inner or outer circle of the ring can start the rotation of this ring. Grabbing of any connection (L_Press) will move the whole set of rings. There are several limitations on the personal movements of nodes (minimum radius for the inner border of the smallest ring, minimum distance between rings, minimum width of the ring), which will certainly affect the `MoveContourPoint()` method.

In the corner of the form there is a panel (moveable!) where several controls will allow you to add or delete rings, change the number of sectors in any ring and renew the values for any ring.

`DefineContour()` method is really simple, and several remarks will help you to undesrtand its code:

- nodes are calculated beginning from the inner ring;
- each new pair of nodes represent the line on the border of two sectors;
- all even nodes are on the inner borders of the rings (odd nodes – on the outer borders).

Moving of the whole object (`Move()` method) means moving of the centers of all the rings and the title.
```
public override void Move (int cx, int cy)
{
    Size size = new Size (cx, cy);
    ptCenter += size;
    foreach (DoughnutRing ring in rings)
    {
        ring .Center = ptCenter;
    }
    title .Move (cx, cy);
}
```

`MoveContourPoint()` for `DoughnutSet` class is definitely more complicated than anything we were looking at before, as for forward movement of any node I have to take into consideration all the previously mentioned limitations. By forward movement of the node I mean the movement along its radius, that will change the width of the ring. The code for forward movement is a bit long to include it here, but several words of explanation will make the undesrtanding of this code easy. For such individual movements of nodes I have first to check (calculate) to which ring and which side of the ring this node belongs; for inner nodes of the rings there is special situation for inner ring and for outer borders there is a special situation for the biggest ring.

For rotation the code is much shorter as all the rings are working under the same rule
```
else if (catcher == MouseButtons .Right)
{
    // Rotation
    double angleNew_Radian = -Math .Atan2 (ptM.Y - ptCenter.Y, ptM.X - ptCenter.X);
    double angleNew_Deg = angleNew_Radian * 180 / Math .PI;
    if (Math .Abs (angleNew_Deg - fStart_Deg [iSectorInRing]) > 1.0)
    {
        int nAddAngle = Convert .ToInt32 (angleNew_Deg -
                                            fStart_Deg [iSectorInRing]);
        rings [jRing] .StartingAngle =
                Auxi_Common .ChangedAngle (rings [jRing] .StartingAngle, nAddAngle);
```

```
        DefineContour ();
    }
}
```

I mentioned before that for rotation I found it more accurate to base calculations on the mouse position and not on shifts, and you can see from this code that I am using mouse position ptM; in the longer part of the same method for forward movement the calculations are based on shifts `cx` and `cy`.

If you expect that for this not simple object you would have to write more complicated code for those three mouse events, you would be disappointed. On the contrary, the code can not be simpler than here; if there was any complexity for this case, then it was accomplished in `DefineContour()` method.

```
private void OnMouseDown (object sender, MouseEventArgs mea)
{
    Movers .CatchMover (mea .Location, mea .Button);
}
private void OnMouseUp (object sender, MouseEventArgs mea)
{
    Movers .ReleaseMover ();
}
private void OnMouseMove (object sender, MouseEventArgs mea)
{
    Movers .MovingMover (mea .Location);
    if (Movers .MoverCaught)
    {
        Invalidate ();
    }
}
```

Pay attention that because you are working with both types of movements, you have to pass the pressed button as a second parameter to `Movers.CatchMover()`.

## Another branch of evolution

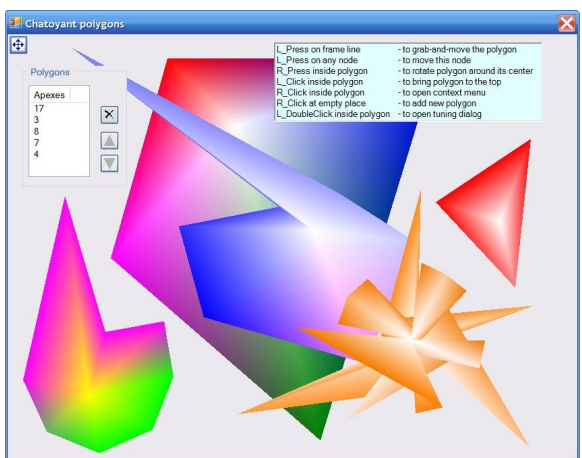
Figure 5. Chatoyant polygons

Throughout all the previous explantion I tried my best to show that in order to turn the existing class of objects into moveable and resizable or in order to design the new class with such features this class must be derived from `GraphicalObject` and override three methods of the last class. Now I am going to show that there can be another way with the same excellent results; look at this way as on another branch of evolution.

C# does not support multiple inheritance of classes; in case you really need it, there are interfaces that will help you to go around this limitation. But if you have spent a lot of time on design of your system of classes, and the system is working perfectly in your applications, the requirement of mandatory inheritance from `GraphicalObject` in order to implement moveable / resizable graphics will not inspire you at all. Here is the way to help you in such a case.

In **TuneableGraphics** application there is one form, which you can fill with the unlimited number of colored chatoyant polygons. You can move them around and reconfigure in any possible way; you can switch on their contours and see that they consist of standard nodes and connections. On initialization each object is a regular polygon; nodes are in the apexes and in center point. All nodes, including the center point, can be moved anywhere, producing very strange figures. Rotation is always around the center point, but for rotation you can R_Press anywhere inside the figure (and not only on other nodes). The polygons look and behave like all other moveable and resizable objects that I was writing about. "If it looks like a duck, quacks like a duck and flies like a duck, then it must be a duck." Usually it is right, but not in this case: `ChatoyantPolygon` is not derived from `GraphicalObject`.

```
public class ChatoyantPolygon
{
    Graph graph;
```

```
    …
    public Graph GetGraph
    {
        get { return (graph); }
    }
```

To implement the whole moving / resizing algorithm `ChatoyantPolygon` includes a field of the `Graph` class, which is derived from `GraphicalObject`. All three mandatory methods for this class are very simple. The object of this class is simply a set of points with the connections between them. `Move()` for this class means moving every point of the set. `MoveContourPoint()` is at the same level of simplicity, as there are no restrictions from the neighbours; any point can be simply moved and that's all.

The objects of `ChatoyantPolygon` class looks interesting and they can be transformed into really complicated figures, but from the point of involving in moving and resizing they differ from all previously mentioned graphical objects only into the direction of simplicity. Paradox? All previously mentioned graphical objects had some basic primitive forms (`Rectangle`, `Point`) and some sizes; combination of these values gave us initial shape to which we added the contour. While organizing those classes we had to write in `Move()` some simple, but needed code for movements of those primitive forms and in `MoveContourPoint()` take into consideration all possible restrictions from the neighbouring nodes. In `ChatoyantPolygon` there is simply nothing except `Graph`, so after writing the code for this class (and I mentioned that it is extremely primitive) we don't need to add anything to `ChatoyantPolygon`.

The initialization of `ChatoyantPolygon` includes the initialization of `graph`, thus calling its method `DefineContour()`. Class `ChatoyantPolygon` has the property `GetGraph`, which is returning this designed `graph`; `Mover` will be dealing with the contour of this `graph`.

Let's decide that we have organized an object of `ChatoyantPolygon` class
`ChatoyantPolygon chatPoly = new ChatoyantPolygon (…);`

We can't register this object as moveable / resizable and add it to `Mover`'s list, because the object itself is not derived from `GraphicalObject`, but we can try to use its `graph`

`Movers .Add (new MovableObject (chatPoly .GetGraph));`

The same three mouse events will be used and in absolutely the same way as with all other objects; there is only minor addition for the case of rotation. On starting the rotation (and it can be started by R_Press at any inner point of the object) I remember the initial angle of the mouse and of all the nodes. On any new mouse position I know the change of its angle, use it to adjust the angles of all nodes, and the whole polygon of any shape is rotating with the mouse. There is a catch: the `Graph` class has such a primitive (simple) `MoveContourPoint()` that there is absolutely nothing for the case of rotation. So in case, when the mouse cursor is rotated and all nodes are adjusted to its turn, I get the new (correct!) graph from rotated polygon and substitute the existing element in `Mover`'s list with the new one.
`Movers .RemoveAt (iPoly);`
`Movers .Insert (iPoly, new MovableObject (chatPoly .GetGraph));`

I can't say that the classical way of organizing moveable / resizable graphical objects by deriving them from `GraphicalObject` is better or worse than this one. When I am designing the new classes, I use the classical way, but I don't see any negative sides in this one also. So the decision on one way or another is up to you.

### And controls also

All the code samples that I showed before and all the explanations were about the design of moveable / resizable graphical objects and working with such objects. But a lot of applications are based on coexistence of graphical objects and controls, and if the mechanism of making graphics alive will simply ignore controls, then for areas of such applications the developed algorithm will lose a significant part of its value. I started the whole design under the idea of eliminating the difference between graphical objects and controls (that are moveable and resizable by default), so everything was designed under the main idea that it has to work identically with both types of elements. The mechanism of turning graphical objects into moveable and resizable is working also with the controls; you can see the moveable panels in nearly each form of **Test_MoveGraphLibrary.** However, there is a small difference in organizing contours for these two types of elements.

The contour for graphical object is organized at the moment of object's initialization in `DefineContour()` method. Two other methods, which are describing the movement of an object as a whole (`Move ()`) and separate movements of each node (`MoveContourPoint ()`), must be also developed for each class of graphical objects, which are going to be moveable and resizable. All three methods are used by `Mover` from the moment this graphical object is included into `Mover`'s list.

Controls do not have such methods but they can be also included into the list of moveable objects

```
Movers .Add (panel);
```

At this moment the `Mover` itself will organize the contour for this control. By default each control gets nonresizable contour slightly out of its boundaries, and from this moment this control is moveable exactly in the same way as any graphical object. Moveable, and in this case not resizable. But any control is resizable by default; can we use this feature of all controls? Certainly, there is another method for making control really resizable by giving it more flexible contour.

```
Mover .Add (Control control, ContourResize resize,
            int minW, int maxW, int minH, int maxH)
```

The second parameter defines the possibilities in changing of the contour's sizes, and the changes in contour will be easily transferred into the resizing of the control itself. I hope that the names of the constants in this enumerator list are clearly explaining each case.

```
enum ContourResize { None, NS, WE, Any };
```

The last four parameters of organizing control's resizable contour define the ranges of possible changes in width and height of the control.

Controls can be turned into resizable individually; I prefer to add a contour to panel or GroupBox thus giving one contour to the group of controls, which I want always to keep together. Relocation of controls on the panel is easily organized through the standard `ClientSizeChange` event of this panel; you can see an example of such panel and associated code in **Form_MovablePlots.cs**. The possibility of moving groups of controls around the screen is so valuable, that I am using it widely enough.

### Moveable graphics began from the standard plotting

I wrote at the beginning that the work on design of moveable / resizable graphics was triggered by the huge demand for such graphics in the areas of scientific and engineering applications. The results turned out to be extremely interesting for areas far away from the original, but for this particular area the new results are not simply some kind of improvement, they have to become a revolution in design of the most complicated applications. In the **Test_MoveGraphLibrary** application there are three different forms that demonstrate different levels of adding new features to standard plotting; these three forms are available via submenu positions under *Y(x) plots*. All these forms are using the `FullGrArea` class, which was specially designed for the plotting of Y(x) functions and parametric functions; this class is included into **MoveGraphLibrary.dll**. You can find the detailed description of `FullGrArea` class together with the tuning forms, which are also provided, in the file **MoveGraphLibrary_Graphics.doc**. Now let's look at the different steps of turning unmoveable standard plots into fully moveable and resizable.

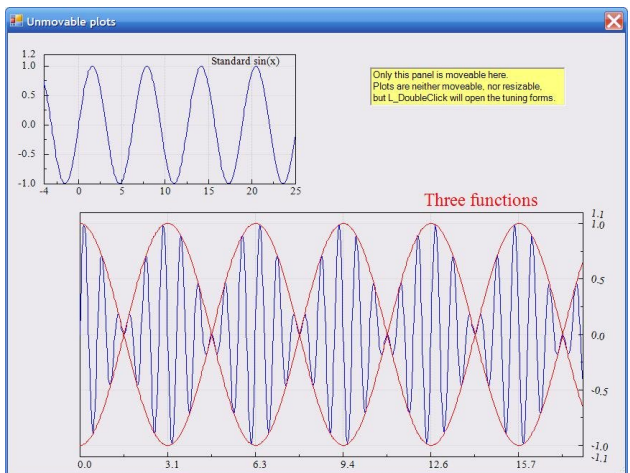

Figure 6. Unmoveable plots

The first form in this group is **Form_UnmovablePlots.cs** with two standard unmoveable graphical objects. These two objects of `FullGrArea` class can be easily turned into moveable / resizable, as any object of `FullGraArea` class has these features. But these two objects were not added to the list of `Mover`, and that was all; they can't move by themselves, somebody has to look after them. What is interesting, there is a `Mover` in this form, and it is working: the colored information panel in the form is moveable. If you want to eliminate all movements of this panel in order to receive the standard form of our days, you can do it easily: comment whatever is dealing with `Movers`. This includes `OnMouseDown()` and `OnMouseUp()` methods and the second part of `OnMouseMove()` method. The first part of the last method is changing the mouse cursor to inform about the possible L_DoubleClick for opening of the tuning forms. This tuning mechanism is working regardless of whether the `FullGrArea` object is really moveable at the moment or not; you can check that in `OnMouseDoubleClick()` method `Movers` is not mentioned.

The second form of this group – **Form_OneMovablePlot.cs** – demonstrates, that in order to turn the unmoveable plot into moveable you need to add a single line of code

```
Movers .Add (area);
```

Everything else in this form is absolutely the same, as in a previous one (well, one plot instead of two). It is nice to have moveable, resizable and tuneable plotting, but still the question of what to show here is still decided by the designer; I decided to show one plot, and it is impossible to see anything else without rewriting the code.

The real power of moveable / resizable graphics for the area of scientific and engineering applications will be unleashed, when users will have a chance to decide **what**, **when** and **how** to show. This is the case of **Form_MovablePlots.cs**.

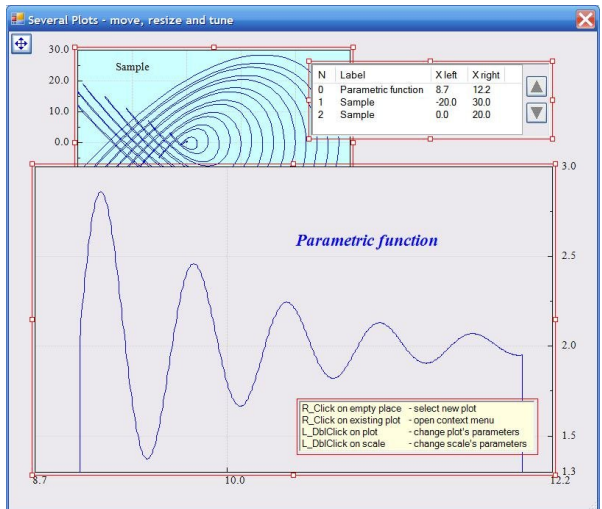

Figure 7. Moveable plots and panels

For any other form of the **Test_MoveGraphLibrary** application I can easily show the picture and write that "this is how it will look". For any, but this. Here you can organize any number of different plots, all of them are moveable and resizable, so I can't predict, how this form will look. But as a designer I have to provide the instrument, with which you can analyze different functions. In this application there is a small limitation, that you can select the functions only from predefined set; in the main form of **TuneableGraphics** application exactly the same instrument is realized without such limitation; there you can explore any number of arbitrary functions.

That's the main idea of applying the new moveable and resizable graphics to the area of scientific and engineering applications: the new technique gives a chance to turn the powerful calculators (our day programs) into real research instruments.

From programming point of view the implementation of moveable / resizable plotting for unlimited number of areas only slightly differs from the previous case of single plotting area. Here we have the same single `Mover` and the same very simple methods for `MouseDown`, `MouseUp` and `MouseMove` events. To organize without any problems (and mistakes) moving, visualization and tuning, I had to organize some reliable link between the function and the area, where it is shown. This is done with the help of additional class `PlotOnScreen`

```
public class PlotOnScreen
{
    DemoFuncType funcType;
    FullGrArea grArea;
```

Functions, which you would decide to show on the screen, are organized into the list
```
List<PlotOnScreen> plotInView = new List<PlotOnScreen> ();
```

All possible operations – adding functions, painting, moving, tuning, changing the order, deleting – all of them are organized only with the objects from this list, so there will be no discrepancies between the real function and its view.

There are three plots (one is out of view) and two panels on this figure; all the plots are certainly moveable and resizable. Panels are also moveable, but only one of them with the list of functions' information is resizable; you can see the difference in their contours.

I have a huge experience in design of very complicated scientific and engineering applications, and I understand very well that turning of such applications from designer-driven into really *user-driven* is not a five minute work. The simple change of graphics, which will turn all plots into moveable / resizable, without any other changes in design of application can be done really quickly. The development of programs inside the new paradigm of *user-driven applications* would require some thoughts from the designers. But users' benefits from the new applications will be enormous.

## Any new problems from the new features?

I have mentioned several times that adding new features to the existing graphical objects can be the cause of some conflicts; they are the same kind of problems that occur on adding new buildings into populated urban area. The areas of our applications is well populated; we have limited number of available mouse generated events, each event is strongly (officially) linked with some expected reaction, and into this organized universe I am trying to add something absolutely new. Certainly, there would be conflicts.

I tried to eliminate or minimize these conflicts by careful design of contours and by shrinking the area of new commands to the proposed system of nodes and connections. However, there can be different ways of solving the conflicts between old and new. I want to mention here some of the rules, that I tried not to break, together with some thoughts for and against my decisions; may be you'll come to something better, when you'll start using moveable / resizable graphics.

- I am using L_Press to start forward movement of the objects or reconfiguring; I use R_Press to start rotation.

- Forward movement of an object is started by L_Press only in the sensitive area of any connection. Certainly, it's not a problem to start such movement by pressing anywhere inside the object, but here is the first conflict: L_Click is widely used to change the order of objects and bring the touched one on top of all others. I found it a very useful feature that in current version of moveable graphics and with implemented system of commands I can grab and slightly move one plot somewhere underneath just to see its part without bringing it on top and destroying the whole view of the form. If the starting of forward movement will be transferred from the nodes' area to the whole area of the object, then the object in move will be automatically brought to the top position. That is exactly how Windows is doing now on the upper level, so it is one of the possible decisions, but still I think that possibility of moving some under layer without disturbing the whole hierarchy is very useful.

- Rotation of the graphical object is usually started by R_Press inside the node. However, it is fairly easy to organize such start by pressing at any point inside the graphical object; for example, in the **TuneableGraphics** application I am showing the sample of absolutely arbitrary polygons (not convex, but any) and the rotation of those polygons can be started at any point inside the object. Starting the rotation by R_Press anywhere inside the object would be much easier from users' point of view, and the reason for not implementing this technique for `PieChart`, `Ring` and so on is simple – I am not sure that it would be correct. Usually the right mouse click is used to pop-up the context menu and I don't like the idea of starting two different processes with the same mouse click. And if the starting of rotation will be expanded to the whole object's area and starting of the forward movement will be still restricted to the area of connections, this will look really awkward.

I am introducing the absolutely new moveable / resizable graphics. I think that this type of graphics will spread quickly enough (the advantages of such graphics is impossible to overestimate), but the development of applications on such graphics will require some setting of commonly used rules on moving and rotation of the objects. It will be much better for users if all the applications will implement movement / resizing based on the same commands.

## Final remarks and some ideas

Moveable / resizable graphics is a very powerful instrument. By adding these features to the objects inside your program you are receiving on the inner level the same flexibility that windows provide on the upper level, and it's not surprising at all that the result is very similar to what we have on the upper level. On the upper level all windows are included into Z-order and the last one to be called is shown on top of all previous windows; when you click any window the system changes the Z-order and move the touched one to the top. Now you have an instrument to do exactly the same at the inner level; the only difference that the system is not going to do it somewhere behind the curtains; you, as a designer, is going to play this role, and there are all the instruments (methods) to do it easily.

The idea of moveable / resizable graphics is not simply new; it is so revolutionary, that even excellent programmers begin to understand the novelty of its invention somewhere from the third explanation. (Up till now I met only one person, who understood the level of this invention in two minutes, but the person is an outstanding one.) Even if you are working with the new things all the time, you are ready to understand and estimate the new results if they are 10-15, may be 20 percent farther on, than the current level. If they differ from your expectations for 100 percent, it is really difficult to accept. The pure psychology of human nature.

It's very difficult to predict, what moveable / resizable graphics can bring into all the existing applications. Extrapolation is used as a one way instrument. How about trying to look back and trying to imagine our PC world with the Windows system stripped of its flexibility? Can you imagine the Windows in which not you but the system is deciding what and where to show? That's what we have now on the inner level. I am absolutely sure, that after spreading of moveable / resizable graphics our modern day fixed applications would be looked at as DOS applications today.

**Test_MoveGraphLibrary** application was designed to show in details this new technique of turning graphical objects (and controls) into moveable and resizable. To make thing absolutely clear I tried to keep all sample classes and all forms for demonstration as simple as possible, that's why each form is working with only one particular moveable / resizable class of objects. However, in real applications you would often like to use together different moveable objects, and to show that there is no problem in doing it I put into **Form_Main.cs** the objects of absolutely different types that are used for explanations in other parts of this application. It's really strange to see in one form standard scientific plotting y(x), houses, set of colored rings and a couple of colored polygons. They have nothing in common except that their design is based on the same described algorithm and they can all live in the same space (form) without any conflicts.

The idea of easy turning of all graphics into moveable / resizable is absolutely new. Though even the whole mechanism of design of new objects with such features was made very simple, and the work with such objects is even easier, still it is an absolutely new thing. For better understanding of what this thing will allow look into **TuneableGraphics** application, that includes a lot of different samples. The idea of that application was not to develop something extraordinary for one particular area; the main idea was to demonstrate the possibility of amazing results for absolutely different objects and areas.

May be the area that can benefit most of all from using of moveable / resizable graphics will be the "financial" graphics. Millions of people are analyzing the financial data and the effectiveness of their analysis greatly depends on how close the view of all these numerous plots is to their expectation of the best and the most informative view. Nothing can be better than the system that allows everyone to look on the plots with the best view personally for him (or her). When you are analyzing a lot of information on the screen and at any moment without stopping the application and going somewhere else for extra tuning you can simply rearrange the whole view, resize and put side by side the pieces you need to compare visually just at this moment – nothing can be better than such system.

The area of scientific and engineering applications, which triggered my work on moveable / resizable graphics, can not only benefit on new type of plotting, but the most complicated systems, by using this new plotting, can be redesigned, turned from designer-driven into *user driven applications* and bring the analysis of the most difficult engineering and scientific problems to another level. Instead of developing several scenarios for calculations up to tiny details, programmers will have to design instruments, which will allow any possible scenario or variant of research work in one or another particular engineering / scientific area. I hope that based on this new technique the engineering / scientific applications can be turned more from being the big powerful calculators into the research instruments.

There can be other areas, which will benefit from using this new moveable / resizable graphics. I designed a very powerful and flexible instrument; specialists in different areas will find the best way to use this new instrument for their applications.

**Acknowledgments**

Many thanks to those with whom I was discussing for years the design of scientific plotting. Special thanks to Dr. Stanislav Shabunya from the Heat and Mass Transfer institute. My work on tuneable graphics was triggered by our mutual project years ago and some significant improvements in clarity of this presentation were the results of his remarks during the last months.

**References**
1. Hawking, S. The universe in a nutshell. Bantam Press 2001.
2. Charles Perrault "Cendrillon".

Dr. Sergey Andreyev ( andreyev_sergey@yahoo.com )
September 2007

## Programs and documents

**Design_of_MoveableResizableGraphics.doc** is one part of several programs and documents that are designed for using of moveable and resizable graphics. The whole package can be divided into several major parts:

| | |
|---|---|
| **MoveGraphLibrary.dll** | the library. |
| **MoveGraphLibrary_Classes.doc** | description of classes included into this library. |
| **MoveGraphLibrary_Graphics.doc** | description of plotting, implemented in this library, and description of all tuning dialogs. |
| **TuneableGraphics.exe** | this application demonstrates moveable / resizable objects from absolutely different areas. |
| **TuneableGraphics_Description.doc** | description of this program. |
| **Test_MoveGraphLibrary.zip** | contains the whole project, demonstrating the use of MoveGraphLibrary.dll for design and use of different moveable / resizable graphical objects. |
| **Design_of_MoveableResizableGraphics.doc** | introduction to the design of moveable / resizable graphical objects; it is the current document. |
| **NewDesignParadigm.doc** | an article about the ideas and consequences of using such graphics in complicated programs. |

All files are available at www.SourceForge.net in the project MoveableGraphics. Look for the latest version.